\documentclass[
]{ceurart}

\usepackage{listings}
\usepackage{wrapfig}
\usepackage{svg}
\usepackage{xparse}

\begin{document}

\copyrightyear{2026}
\copyrightclause{Copyright for this paper by its authors.
  Use permitted under Creative Commons License Attribution 4.0
  International (CC BY 4.0).}

\conference{}

\title{From Multimodal Signals to Adaptive XR Experiences for De-escalation Training}

    \author[1]{Birgit Nierula}[%
    orcid=0000-0003-2174-5872,
    email=Birgit.Nierula@hhi.fraunhofer.de,
    ]
    \cormark[1]
    \fnmark[1]
    
    \author[1]{Karam Tomotaki-Dawoud}[%
    orcid=0009-0006-2745-8312,
    ]
    \fnmark[1]
    
    \author[1]{Daniel Johannes Meyer}[%
    orcid=0009-0009-6039-7234,
    ]
    \fnmark[1]
    \address[1]{Department for Vision and Imaging Technologies, Fraunhofer Heinrich-Hertz-Institute, Einsteinufer 37, 10587 Berlin, Germany}

    \author[1]{Iryna Ignatieva}
    
    \author[1]{Mina Mottahedin}
    
    \author[1]{Thomas Koch}
    
    \author[1]{Sebastian Bosse}
    \cortext[1]{Corresponding author.}
    \fntext[1]{These authors contributed equally.}

\begin{abstract}
We present the early-stage design and implementation of a multimodal, real-time communication analysis system intended as a foundational interaction layer for adaptive VR training. The system integrates five parallel processing streams: (1) verbal and prosodic speech analysis, (2) skeletal gesture recognition from multi-view RGB cameras, (3) multimodal affective analysis combining lower-face video with upper-face facial EMG, (4) EEG-based mental state decoding, and (5) physiological arousal estimation from skin conductance, heart activity, and proxemic behavior. All signals are synchronized via Lab Streaming Layer to enable temporally aligned, continuous assessments of users’ conscious and unconscious communication cues.

Building on concepts from social semiotics and symbolic interactionism, we introduce an interpretation layer that links low-level signal representations to interactional constructs such as escalation and de-escalation. This layer is informed by domain knowledge from police instructors and lay participants, grounding system responses in realistic conflict scenarios.

We demonstrate the feasibility and limitations of automated cue extraction in an XR-based de-escalation training project for law enforcement, reporting preliminary results for gesture recognition, emotion recognition under HMD occlusion, verbal assessment, mental state decoding, and physiological arousal. Our findings highlight the value of multi-view sensing and multimodal fusion for overcoming occlusion and viewpoint challenges, while underscoring that fusion and feedback must be treated as design problems rather than purely technical ones. The work contributes design resources and empirical insights for shaping human–AI-powered XR training in complex interpersonal settings.
\end{abstract}


\maketitle

\section{Introduction}
Human communication is inherently multimodal, combining verbal content with non-verbal signals such as body posture, gestures, facial expressions, vocal prosody, and interpersonal space~\cite{jewitt2014routledge}. In emotionally charged interactions, these non-verbal cues often carry as much -or more- meaning than spoken language~\cite{Otu_2023}. They are frequently produced unconsciously~\cite{lakin2006automatic}, yet they strongly shape how intentions are perceived and whether an interaction escalates or de-escalates~\cite{Otu_2023}. For professionals operating in high-stakes contexts, such as law enforcement~\cite{rozelle1978interpretation} or healthcare~\cite{schmidmast2013nonverbal}, the ability to regulate both verbal and non-verbal communication is therefore a critical skill. 

Traditional communication training methods~\cite{white2023_Evaluation_TempleDeescalationTraining, sjoberg2024_policeEducation, gelis2020_rolePlay}, including role-play and instructor-led feedback, can support the development of these skills but are limited by subjectivity, delayed feedback, and the difficulty of making unconscious communication patterns visible to trainees. Video-based review and post-hoc analytics offer additional insight, but they separate reflection from the lived, embodied experience of the interaction itself.

Immersive virtual reality (VR) provides a promising medium for communication training by enabling embodied, situated interactions that evoke realistic emotional and behavioural responses while remaining safe, repeatable, and controllable~\cite{sanchez2005presence}. Prior studies have shown that VR can elicit affective and physiological reactions comparable to real-world encounters~\cite{tootell2021_proxemics,marin-morales2021_emotional_arousal}. However, designing VR systems for interpersonal skill training introduces a fundamental challenge: communication unfolds continuously across multiple modalities, and its meaning is highly context-dependent~\cite{hodge1998social}. Systems that rely on predefined choices, scripted branching, or post-session evaluation therefore struggle to capture the dynamic nature of real-world interactions.

To shape meaningful training experiences, XR environments must be able to sense and interpret users’ communicative behaviour in real time and translate it into adaptive responses that remain immersive and interpretable. This raises a central design question for XR: how can multimodal communication signals be operationalized in a way that supports adaptive, experience-driven training without disrupting the interaction itself? Addressing this question requires not only accurate signal processing, but also careful design decisions regarding which cues matter, how they are interpreted, and how they influence virtual character behaviour and feedback.

\paragraph{Research Gap.} Limitations of current solutions are their lack of objective feedback and scalability as they are typically implemented as role-play and instructor-led feedback. Objective and scalable XR solutions that allow to capture and predict the dynamics of social interactions are challenged by a lack of multimodal (verbal and non-verbal) communication sensing in real-time in a context-aware  manner. This, however, is necessary for adaptive, context-aware avatar responses that combined with objective and transparent feedback can foster awareness and control over own verbal and non-verbal communication cues. Implementing this requires both technical signal processing capabilities and social theory design frameworks for operationalizing communication cues.

\paragraph{Contributions.}
This work offers the following contributions to the design of multimodal communication sensing XR systems grounded in social science:

\begin{itemize}
    \item \textbf{Implemented and evaluated:}
    \begin{itemize}
        \item A multi-view skeletal gesture recognition pipeline achieving 82--88\% accuracy for conflict-relevant postures and gestures;
        \item A late-fusion architecture combining lower-face video with upper-face EMG for emotion recognition under HMD occlusion, achieving 51\% macro-F1 on 7-class classification;
        \item Physiological arousal analysis (SCR, HRV) validated in a controlled experimental scenario with sensitivity to proxemic violations and emotional facial expressions;
        \item Verbal communication analysis including formality detection (68\% accuracy)
        \item prosodic feature extraction with preliminary correlation analysis.
    \end{itemize}
    
    \item \textbf{Implemented with preliminary evaluation:}
    \begin{itemize}
        \item EEG-based mental state decoding using SSD+SPoC decomposition, demonstrated as proof-of-concept in 3~participants;
        \item Sentence complexity assessment, evaluated on external benchmark data but not yet validated in police interaction contexts.
    \end{itemize}
    
    \item \textbf{Conceptual contributions:}
    \begin{itemize}
        \item An integrated system architecture synchronizing five parallel analysis streams via Lab Streaming Layer for temporally aligned multimodal assessment;
        \item A design framework grounded in social semiotics and symbolic interactionism, linking low-level signal representations to interactional constructs (escalation/de-escalation) through domain expert and lay participant input;
        \item Design considerations for multimodal fusion and adaptive feedback in XR training, framed as open design problems rather than solved technical challenges.
    \end{itemize}
\end{itemize}

\noindent By distinguishing evaluated components from preliminary and conceptual contributions, we aim to provide transparency about the current maturity of each system element.

We illustrate this approach through its application in an ongoing research project on de-escalation training for law enforcement. We report preliminary findings and design reflections that suggest the feasibility of using multimodal communication analysis to inform real-time adaptation in immersive environments. By foregrounding design considerations and early insights, this work contributes to the ShapeXR community by highlighting how human–AI-powered XR experiences can be used to support embodied learning in complex interpersonal scenarios.

\section{Related Work} 

In recent years, a range of product developments and research activities have sought to exploit the advantages of XR-based training in the policing context. The TARGET innovation project~\cite{target_toolkit} developed a toolkit for creating integrated AR/VR training scenarios, which also enables the representation of complex, pan-European large-scale operations. Similarly, many existing training solutions focus on scenario simulation to train procedures, behavioural patterns, decision-making, and communication from a tactical, process-oriented perspective~\cite{relion_website, refense_website, streetsmart_website, axon_website}. These approaches often incorporate haptic elements through the use of operational equipment, particularly weapons~\cite{operatorxr_website}. Data-driven tactical and behavioural analyses are typically conducted after the training, based on video data, for example~\cite{holoforce_website}. In contrast to these existing trainings, our approach does not aim to model the operational process as a whole, but instead focuses on communication dynamics in critical situations with the aim to avoid the use of operational equipment, particularly weapons.

In the domain of communication training, both commercial products and research contributions demonstrate the benefits of data-driven communication analysis. For instance, VR trainings for presentation training incorporate speech-related quality metrics to provide automated post-training evaluations~\cite{vreasyspeech2024}. In the context of police training, research shows the usefulness of VR communication training for critical situations, as well as the benefit of and demand for evaluation and visualization of the training~\cite{murtinger_2024}~\cite{munoz2024psychophysiological}.
As an interactional feedback mechanic, in 2013 a job interview simulator~\cite{job-interview} leveraged the recognition of nonverbal communication cues~\cite{hutchison_nova_2013} not only to generate useful feedback but also to model specific communication dynamics within training scenarios in real-time. While existing studies have established the potential value of such approaches in principle~\cite{bartyzel_2025}, interactive feedback typically focuses on specific aspects of communication. To allow progress towards a more generalizable model of communication, the presented approach develops a system in which a broad range of relevant modalities, combined with potentially interpretative signal analysis, serves as the foundation for designing a more comprehensive feedback system for VR trainings.

Recent work in applied virtual environments has investigated combining neural and peripheral physiological signals for user state assessment, including simulator sickness~\cite{tauscher2020exploring} and perceived face realness~\cite{chen2024realness}. Real-time multimodal analysis enables affect-adaptive systems; for instance, EEG and peripheral signals have been used to modulate game difficulty~\cite{Chanel2011SMC}. In parallel, facial electromyography (EMG) has emerged as a robust modality under HMD occlusion: wearable platforms such as EmteqPRO integrate upper-face EMG with physiological sensors in VR-compatible form factors~\cite{Gjoreski2021Ubicomp,EmteqLabsPaper}, enabling expression classification~\cite{kiprijanovska2022facial} and continuous valence-arousal monitoring during immersive viewing~\cite{Gjoreski2022SciRep}. Complementary optical approaches address occlusion through peri-ocular IR imaging~\cite{Hickson2019WACV}, photo-reflective sensors~\cite{Murakami2019SIGGRAPH}, or generative face completion~\cite{Numan2021HMDRemoval}.
Beyond facial cues, recent work highlights the role of body posture and gesture as salient non-verbal communication markers in interactive and conflict-related scenarios~\cite{Tomotaki_24_MVGestRec}. Multimodal fusion strategies have shown consistent improvements over unimodal baselines in VR-based emotion, action, and user-state recognition~\cite{indrasiri2024multimodal,Polo2025CBM}.

\newpage

\section{Extraction Layer}
\begin{wrapfigure}{r}{0.6\textwidth}
    \vspace{-10pt} 
    \centering
    \includegraphics[width=0.6\textwidth]{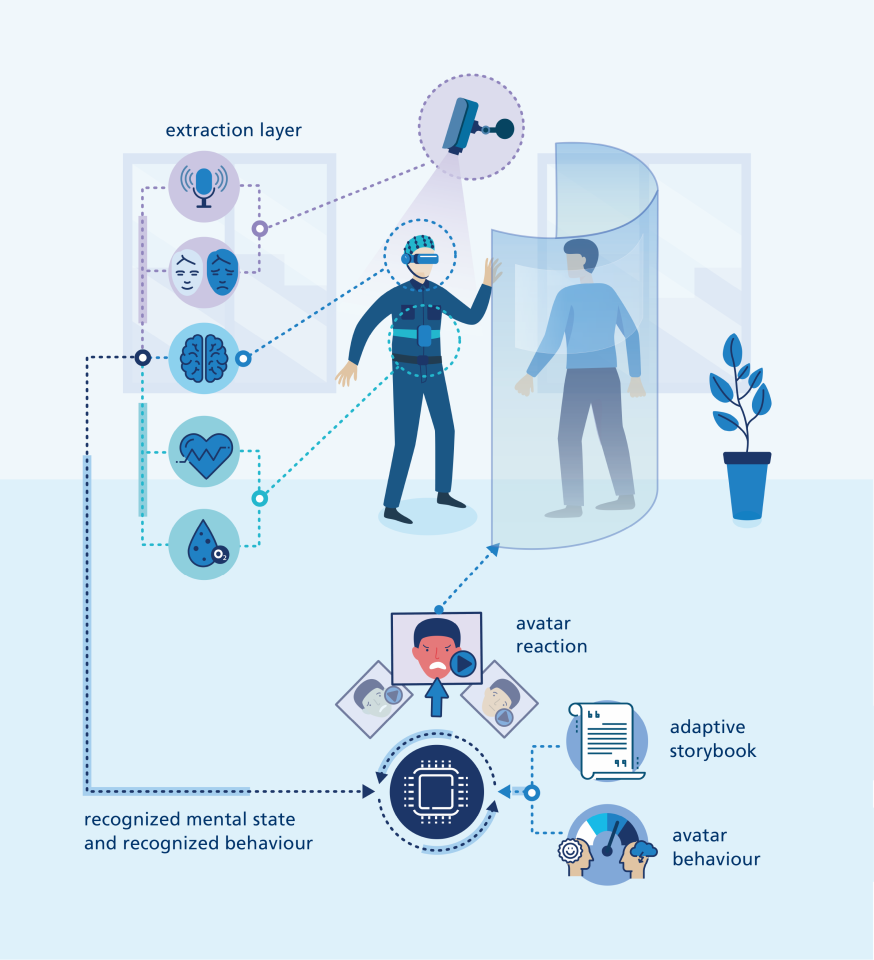}
    \caption{Conceptual system overview of the extraction layer integrated into an adaptive XR experience. The system objectively assesses user experiences during virtual conflicts by extracting features from signal modalities and fusing them with avatar behaviour and storybook context. The resulting signal adapts the virtual experience.}
    \label{fig:k3vrSystem}
    \vspace{-27pt} 
\end{wrapfigure}

We propose an integrated pipeline that extracts communication cues from multimodal sensor data to enable comprehensive feedback in immersive VR training environments.
The system operates on five parallel processing streams: (1) a speech analysis stream including both verbal and prosodic features, (2) a skeletal gesture stream capturing body posture and movement dynamics~\cite{Tomotaki_24_MVGestRec}, (3) a multimodal affective stream integrating facial expressions and physiological signals~\cite{Nierula2026VRemotion}, (4) mental state decoding stream and (5) a bodily arousal stream. 
This architecture enables real-time assessment of both conscious and unconscious communication cues, which can be essential for professional training in domains such as law enforcement de-escalation and healthcare communication.

\subsection{System Overview} 
The pipeline begins with synchronized acquisition from multiple sensors: (i) at least one external microphone, (ii) two or more cameras positioned at complementary angles (e.g., 0° and ±30° relative to the sagittal plane) for capturing body gestures and lower-face expressions, and (iii) a VR-integrated head-mounted display (HMD) equipped with facial electromyography (EMG) sensors on the upper face, 
 (iv) electroencephalography, (v) skin conductance, and (vi) electrocardiography (see Figure~\ref{fig:k3vrSystem} for system overview).

All data streams are synchronized via Lab Streaming Layer (LSL~\cite{kothe2025lab, lsl_github}) with event markers inserted at critical training milestones, ensuring precise temporal alignment.

\subsection{Analysis Streams}
\label{sec:analysis_streams}
\paragraph{Verbal Communication} 

The verbal content of the audio signal is extracted by faster-whisper~\cite{fasterwhisper_github}, an optimized reimplementation of OpenAI’s Whisper model~\cite{radford_robust_2022}. 
Insults are detected by keyword recognition. For complex classification tasks, a transformer-based language model is employed. Our first application focuses on formality detection, since, for example, using the informal “du” instead of “Sie” may be perceived as disrespectful. For this, we used a transformer-based language model (XLM-R-large). The model was fine-tuned on a subset of German data from the FAME-MT~\cite{FAME-MT}  and CoCoA-MT~\cite{CoCoA-MT} datasets to classify user utterances as formal, informal, or neutral.

Given its relevance in the context of language barriers, we explored sentence complexity assessment. For this subtask we followed the approach of Asghari and Hewett (2022)~\cite{Asghari2022} and use both linguistic features and a XLM-R-large model fine-tuned on Text Complexity DE~\cite{Naderi2019} dataset to predict how complex a sentence is on a scale from 1 (least complex) to 7 (most complex).

\paragraph{Nonverbal Speech Parameters}

Closely connected to our perception, prosodic features play an important role in the impact of spoken language.
From the microphone signal, a fixed calibration extracts the absolute sound pressure at a reference distance of 1 meter, allowing for the calculation of psychoacoustic features. Highly connected to the affective content of speech are loudness and pitch~\cite{wu10_interspeech}. Loudness is calculated according to ISO-standard~\cite{iso_loudness} using the MOSQITO-library~\cite{mosqito}. The PYIN algorithm is used~\cite{pyin} to estimate a pitch contour, from which the average pitch as well as minimum and maximum pitch are derived. To extract valence and arousal a speech emotion recognition model is applied~\cite{Wagner_2023}. Based on the transcription of the faster-whisper model, the speaking rate is calculated.

\paragraph{Skeletal Gesture Recognition} 

RGB images from multiple camera views are processed through BlazePose~\cite{Bazarevsky2020BlazePoseOR,grishchenko2022blazeposeghumholisticrealtime, BlazePoseGHUMGHUML}  for real-time 3D pose estimation, extracting body landmarks that serve as kinematic skeletons. Features are derived from pairwise Euclidean distances between 3D joint coordinates, normalized to account for individual anthropometric variation. This representation captures the spatial relationships underlying conflict-relevant gestures (e.g., crossed arms) without dependency on appearance,  where simple neural networks classify these normalized distance features into discrete gesture categories, enabling immediate recognition of escalation or de-escalation signals (according to context described in Sec.~\ref{SemanticInterp}).

\paragraph{Emotional Facial Expressions} 

This stream operates on two complementary modalities addressing the HMD occlusion challenge. The visual branch extracts lower-face features via ResNet-18/50 backbones pre-trained on CK+ with occlusion augmentation (black patches, random noise, VR headset overlays) to focus on mouth and cheek regions. The physiological branch processes seven-channel facial EMG (corrugator supercilii, frontalis, orbicularis oculi, zygomaticus major) using a compact RBF kernel representation that captures inter-muscle correlations within sliding windows. These correlations reflect the complex multi-muscle patterns underlying emotional expressions.

Visual and EMG embeddings are independently projected to 128-D representations and then concatenated into a 256-D fused feature vector, which is passed through fully connected layers to classify seven emotion categories (the six basic emotions introduced by Ekman~\cite{Ekman_1992} plus neutral). This modular late-fusion design enables independent optimization of each stream while allowing seamless incorporation of additional modalities (e.g., skin conductance, heart rate) without architectural changes.

\begin{figure}[t]
    \centering
    \includegraphics[width=\textwidth]{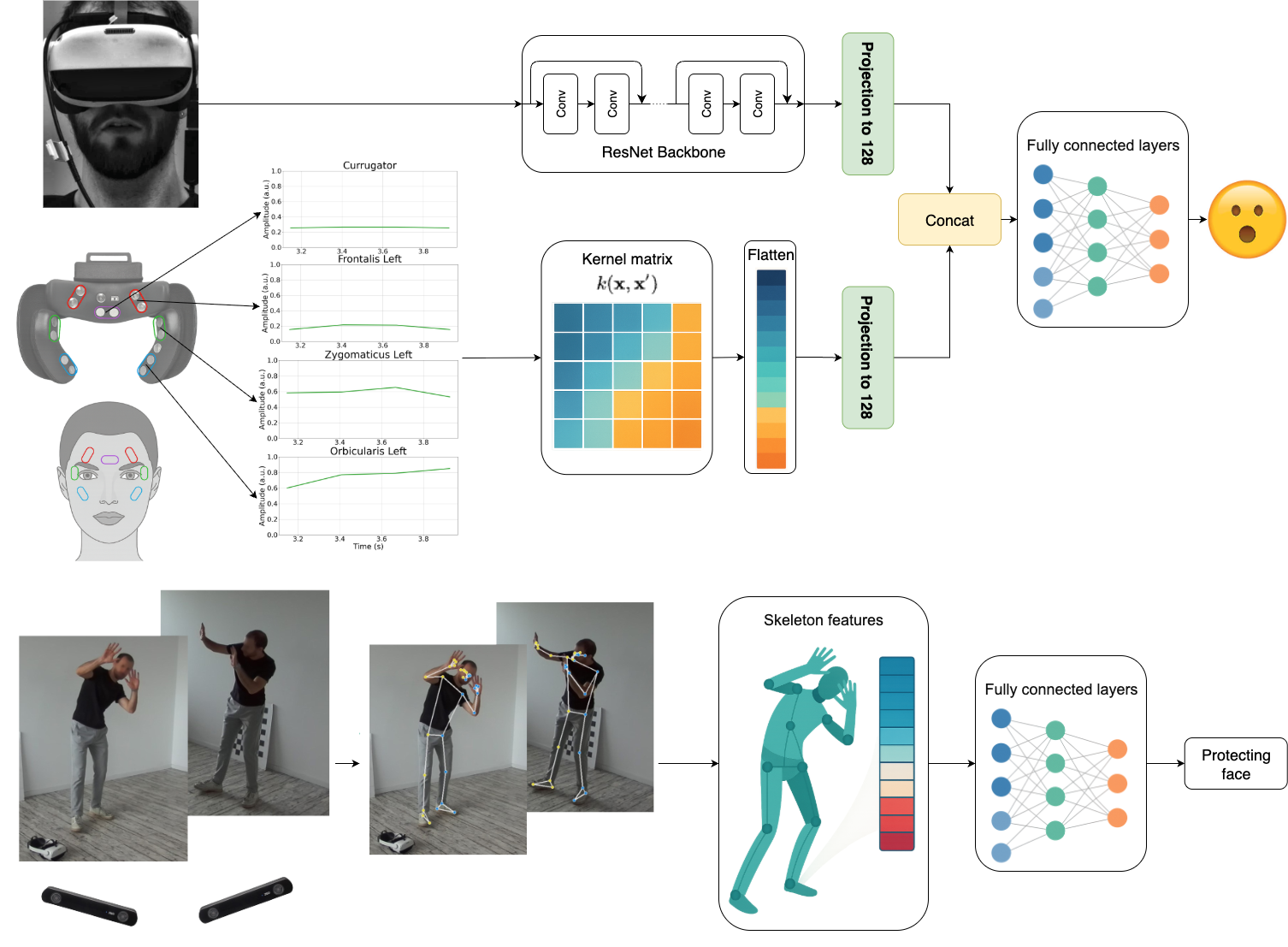}
	\caption{Integrated dual-stream pipeline combining gesture and emotion recognition for VR training feedback. 
	The emotion stream (top) fuses lower-face video with upper-face EMG under HMD occlusion; the gesture stream (bottom) processes multi-view skeletal data to detect conflict-relevant body language. Both streams converge into a unified weighted output that delivers real-time, multimodal feedback for skill refinement.}
    \label{fig:multimodal_pipeline}
\end{figure}

The weighted combination of gesture and emotion recognition streams (shown in fig. \ref{fig:multimodal_pipeline}) enables VR training systems to provide real-time feedback, where gesture recognition raises trainee awareness of unconsciously performed body language, while emotion recognition reflects the emotional impact of their communication, facilitating iterative skill refinement in realistic simulated scenarios. 

\paragraph{Mental State Decoding}

A user's mental state can be extracted via electroencephalography (EEG), a non-invasive technique that measures cortical electrical activity through scalp sensors. To isolate relevant neural signals from the complex mixture of brain activity, we employed multivariate source decomposition methods. These computational approaches enable extraction of specific cortical activity patterns from 64-channel EEG recordings in relation to arousal-inducing stimuli. As a first target variable, we used the avatar's proxemic behavior—specifically, the avatar approaching the user.

Specifically, we employed a two-stage decomposition pipeline combining spatio-spectral decomposition (SSD~\cite{nikulin2011_SSD}) and source power comodulation (SPoC~\cite{daehne2014_SPoC}) to extract arousal-related cortical activity induced by avatar approach. In the first stage, SSD was applied to maximize the signal-to-noise ratio in the alpha frequency band (8-12 Hz), which is known to inversely correlate with arousal states in frontal and parieto-occipital cortical regions. SSD identifies spatial filters that enhance oscillatory components within this target frequency range while simultaneously suppressing activity from flanking frequency bands, thereby isolating spectrally pure alpha oscillations from the multi-channel EEG data. In the second stage, SPoC was applied to the SSD-filtered components to identify neural sources whose alpha power fluctuations showed maximal covariation with the time-varying distance of the approaching avatar. SPoC optimizes spatial filters by finding linear combinations of EEG channels that maximize the correlation between extracted source power and the continuous target variable (avatar proximity). This approach has been shown to identify arousal-sensitive neural patterns in VR users~\cite{hofmann2021_rollercoaster}.

\paragraph{Bodily Arousal Recognition} 
Emotionally charged interactions trigger physiological arousal that can be objectively measured through skin conductance response (SCR) and heart activity. These two continuous signals can capture distinct aspects of the autonomic nervous system response during emotionally charged interactions. SCR reflects sympathetic activation, with higher normalized peak amplitudes indicating greater arousal, while heart rate variability (HRV) reflects parasympathetic activity, with lower values indicating autonomic stress responses. Our analysis pipeline returns (1) an SCR stream providing normalized peak amplitudes and (2) a HRV stream providing the Root Mean Square of Successive Differences (RMSSD) of R-R intervals. The pipeline is described in more detail in~\cite{nierula_2025_physio}.Individual baseline measurements recorded prior to the interaction establish participant-specific thresholds to detect when physiological signals deviate from resting state, indicating arousal.

\paragraph{Proxemic Regulation} 
Proxemic behaviour, the spatial positioning between user and avatar, is extracted from the VR headset and avatar positions. The interpersonal distance reflects how individuals regulate space during emotional interactions and serves as a non-verbal indicator of comfort, threat perception, and social engagement.

\subsection{Illustrative Example}
\label{sec:vignette}

To clarify how the extraction layer informs adaptive behavior, consider the following hypothetical scenario during a de-escalation training session:

A trainee officer attempts to calm a visibly agitated virtual citizen. As the interaction progresses, the system continuously monitors multiple streams (see Table~\ref{tab:vignette_cues}).

\begin{table}[h]
    \centering
    \caption{Example of multimodal cues observed during a hypothetical training interaction.}
    \label{tab:vignette_cues}
    \begin{tabular}{@{}lll@{}}
        \toprule
        \textbf{Stream} & \textbf{Observed Cue} & \textbf{Extracted Feature} \\
        \midrule
        Verbal & Officer uses informal address (``du'') & Formality classifier $\rightarrow$ \textit{informal} \\
        Prosody & Elevated vocal loudness (mean loudness > 8 Sone) & Loudness module $\rightarrow$ \textit{high} \\
        Gesture & Arms crossed, weight shifted backward & Pose classifier $\rightarrow$ \textit{defensive posture} \\
        Physiology & SCR amplitude 1.5$\times$ baseline & Arousal module $\rightarrow$ \textit{elevated} \\
        \bottomrule
    \end{tabular}
\end{table}

These signals are forwarded to the interpretation layer, where context from the scenario storybook (e.g., ``citizen is non-native speaker; prior exchange was tense'') modulates their meaning. The co-occurrence of informal language, elevated loudness, and defensive posture (combined with rising physiological arousal) is interpreted as emerging trainee frustration with potential escalation risk.

Based on this interpretation, the system triggers a conservative avatar response: the virtual citizen takes a small step backward, averts gaze briefly, and lowers vocal intensity; subtle behavioral shifts designed to signal discomfort without abrupt scene changes. A cooldown window (e.g., 5\,seconds) prevents repeated adaptations from transient signal fluctuations. Post-scenario, the trainer reviews flagged moments alongside the multimodal timeline, contextualizing feedback for the trainee.

This example illustrates the intended information flow; however, as discussed in Section~\ref{sec:discussion}, the fusion logic and adaptation policies remain under active development.

\subsection{Method Summary}
Table~\ref{tab:method_summary} provides a compact overview of the core processing parameters for each analysis stream. The extraction layer follows a modular design philosophy that prioritizes methodological transparency while maintaining deployment flexibility.

We intentionally refrain from providing a fixed instrumentation specification (e.g., exact device models, proprietary sensor configurations) because the pipeline is designed to be hardware-agnostic within defined functional constraints. Specific sensors—including cameras, microphones, EEG systems, and physiological recording platforms—can be exchanged depending on deployment context, facility requirements, and evolving technology. For instance, any RGB camera system meeting minimum frame-rate requirements ($\geq$30\,fps) can serve the gesture stream, and various commercial EEG systems supporting 32+ channels and LSL integration are compatible with the mental state decoding pipeline. Similarly, facial EMG can be acquired through standalone electrode arrays or integrated HMD-mounted platforms. 

This flexibility is a deliberate design choice: by decoupling the signal-processing methodology from specific hardware implementations, the system remains applicable across heterogeneous training environments—from laboratory settings to operational training facilities—and can incorporate improved sensors as they become available. The LSL-based synchronization architecture ensures temporal alignment regardless of the specific device combination, provided each stream publishes timestamped data to the shared network.

Minimum functional requirements for each modality are as follows: audio input at $\geq$48\,kHz sampling rate, capable of calibration (e.g. without integrated auto-leveling or noise reduction); RGB video at $\geq$30\,fps with sufficient resolution for pose estimation ($\geq$1080p); physiological signals (ECG, EDA) at $\geq$250\,Hz; facial EMG at $\geq$800\,Hz; and EEG at $\geq$250\,Hz with a minimum of 32 channels for source decomposition methods.

\begin{table}[ht]
    \centering
    \caption{Compact method summary for each analysis stream. Specific hardware can be exchanged within stated functional constraints; see text for minimum requirements.}
    \label{tab:method_summary}
    \resizebox{\textwidth}{!}{%
    \begin{tabular}{@{}p{2.2cm}p{4.2cm}p{3.0cm}p{5.0cm}@{}}
        \toprule
        \textbf{Stream} & \textbf{Preprocessing / Artifact Management} & \textbf{Window / Segmentation} & \textbf{Feature Extraction} \\
        \midrule
        Verbal & High pass (60 Hz); speech-to-text (faster-whisper) & Sentence-wise (based on STT output) & Transformer embeddings (XLM-R-large); keyword matching for insult detection; formality and complexity scores \\
        \addlinespace
        Prosody & High pass (60 Hz); fixed calibration for absolute sound pressure at reference distance of 1 meter; voice activity detection. \textit{For speaking rate:} speech-to-text & Continuous sliding window (e.g. 2.0 s window and 0.5 s hop). \textit{For speaking rate:} sentence-wise separation & Loudness (mean of time-varying loudness according to ISO\,532-1, via MOSQITO); pitch contour (PYIN: mean, min, max); valence/arousal (wav2vec2-based SER); speaking rate (wpm)\\
        \addlinespace
        Gesture & Multi-view frames synchronization and normalization & Per-frame ($\sim$33\,ms at 30\,fps) & 33 BlazePose landmarks $\rightarrow$ pairwise 3D Euclidean distances; normalized feature vector input to Random Forest classifier \\
        \addlinespace
        Facial Emotion (Multimodal) & 
        \textit{EMG:} Notch filter (50\,Hz + harmonics), bandpass 100-400\,Hz; baseline subtraction; amplitude clipping; full-wave rectification; min--max scaling. \textit{Video:} Face detection and cropping; intensity normalization. & 
        
        \textit{EMG:} 1.0\,s window, 0.25\,s hop. \textit{Video:} Single frame per window (temporally aligned with EMG) & 
        
        \textit{EMG:} 7-channel inter-muscle RBF kernel matrix $\rightarrow$ 128-D projection. \newline 
        \textit{Video:} ResNet-50 $\rightarrow$ 128-D projection. \newline 
        \textit{Fusion:} Late fusion; 256-D concatenated embedding $\rightarrow$ FC layers $\rightarrow$ 7-class logits \\
        \addlinespace
        Mental State (EEG) & Bandpass 1--40\,Hz; bad-channel interpolation; ICA-based ocular and muscle artifact removal & 1 second, 50\% overlap; continuous for real-time & SSD $\rightarrow$ alpha-band (individual peak $pm$2\,Hz) enhancement; SPoC $\rightarrow$ components maximally correlated with target variable (avatar distance); alpha power as output feature \\
        \addlinespace
        Bodily Arousal (EDA) & Band-pass filter (0.0159--0.5\,Hz); z-score normalization & Event-locked windows (1--5\,s post-stimulus); continuous phasic monitoring & Phasic SCR amplitude \\
        \addlinespace
        Bodily Arousal (ECG) & z-score transformation; highpass filter $\geq$ 0.5\,Hz; R-peak detection; ectopic and missing peak correction & 5 seconds; no overlap  & RMSSD (root mean square of successive R-R differences) \\
        \addlinespace
        Proxemics & Coordinate transformation to common reference frame & Continuous (real-time, per VR frame) & Euclidean distance between HMD position and avatar centroid; rate of change (approach velocity) \\
        \bottomrule
    \end{tabular}%
    }
\end{table}

\section{From Semantics to Interpretation}
\label{SemanticInterp}
To develop an adaptive system that captures face-to-face interaction in VR, this work draws on insights from social semiotics and symbolic interactionism. Social semiotics provides a framework for decomposing interaction into its multimodal modes and understanding how meaning emerges from their combination~\cite{flewitt2018multimodality}. However, meaning is not solely encoded in semantic units but is also constructed through interactional processes. Symbolic interactionism posits that meaning is produced and continuously reshaped through participants' mutual interpretation during interaction~\cite{blumer1969symbolic}. From this perspective, interactions are inherently co-constructed: participants continuously interpret and adjust their behavior in response to one another's communicative acts.

The design approach pairs detailed semantic analysis of each modality (from social semiotics) with an interactionist perspective that frames escalation and de-escalation as emergent system dynamics. First, modality elements meaningful for system functionality are identified, translating semantic categories into computational representations. Second, the interactional contexts in which these elements occur are captured. Third, the system examines whether these elements are interpreted differently depending on context—reflecting the symbolic interactionist premise that meaning emerges through situated interpretation rather than fixed semantic codes.

This approach is illustrated through a case study on non-verbal cues. Drawing from symbolic interactionism, escalation or de-escalation in police-citizen encounters are often shaped by turning points marked by shifts in nonverbal behavior~\cite{KeesmanWeenink2022, Sunde2024}. Based on literature and interviews with police trainers, 19 gestures were identified and categorized by communicative function (routine/alert, calming, commanding, space-controlling, reactive). A machine learning pipeline enables real-time detection and classification of these gestures (Section~\ref{sec:analysis_streams}). Ten realistic scenarios in which these gestures occur were identified through trainer interviews and provide the context information. An online study with German residents (N=600, stratified by age and gender) then assessed how each gesture could escalate or de-escalate interactions within each context.

This process generates escalation/de-escalation indices for each gesture-scenario pairing, differentiated by officer gender and citizen demographics. These theory-driven mappings operationalize the research framework in three stages: (1) multimodal cues (in the case study gestures) are computationally detected, (2) context-dependent interpretive states (escalation/de-escalation potential) are derived from empirical data reflecting situated meaning-making, and(3) adaptive system responses are triggered—enabling virtual agents to escalate or de-escalate based on trainee behavior, adjusting narrative dynamics according to both character and trainee demographics. The indices thus serve as computational parameters that translate interactionist principles into concrete system adaptations.

\section{Results}

\begin{wrapfigure}{r}{0.4\textwidth}
    \centering
    \includegraphics[width=0.4\textwidth]{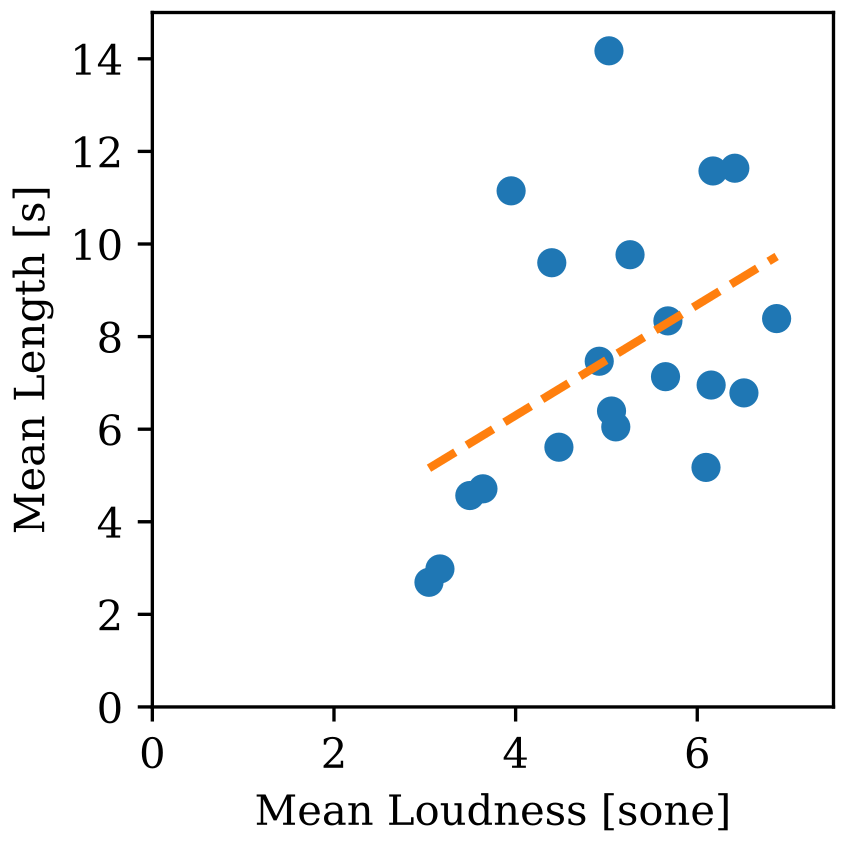}
    \caption{Mean values of length and loudness over answers of N=20 participants with an indicated linear regression.}
    \label{fig:results_prosodic}
    \vspace{-10pt} 
\end{wrapfigure}
Our investigations into automated recognition of communication cues for conflict-related training scenarios yielded promising results across six complementary domains:

For \textbf{verbal communication}, pilot test data was collected from N=20 police officers interacting in a prototypical VR scene for de-escalation training.
Formality classification on the mentioned test set achieved an accuracy of 0.68 and a weighted F1-score of 0.64, indicating limited robustness in capturing formality cues beyond simple lexical patterns. This suggests that the model relies heavily on patterns from the training data and leaves room for improvement through more diverse examples and deeper stylistic features.

For \textbf{nonverbal speech parameters}, a preliminary exploratory data analysis on the pilot test data was performed. Figure~\ref{fig:results_prosodic} shows the mean length and loudness over the answers of each of the 20 police officers. A correlation of $\mathrm{r}=0.46$ was found, suggesting a dependency of these variables. Given that each participant was professionally trained, there still seems to be significant variation. As a combined measure it might be connected to different communication approaches, as for example shorter, quieter answers might hint at a more cautious approach, while louder, longer answers might hint at a more dominant communication.

For \textbf{body gesture recognition}, the multi-view skeleton-based system demonstrated robust performance, with Random Forest classifiers consistently achieving $82–88\%$ accuracy across different feature sets in 5-fold cross-validation. Handcrafted features derived from pairwise 3D joint distances performed comparably to models trained on raw skeleton coordinates, while offering improved interpretability and reduced overfitting. Notably, integrating multi-view camera data enhanced recognition accuracy, particularly in challenging scenarios involving occlusions or non-frontal poses.

For \textbf{facial emotion recognition} under HMD occlusion, conventional image-based methods (e.g., OpenFace3~\cite{hu2025openface}) collapsed to near-chance performance ($\approx16\%$ accuracy on a 7-class prediction) when applied to lower-face-only inputs, highlighting the limitations of transferring existing facial expression recognition pipelines into immersive VR environments. However, our proposed late-fusion architecture—combining lower-face video with seven-channel upper-face EMG—substantially outperformed unimodal baselines, achieving $51\%$ macro-F1 on subject-independent held-out test identities. Unimodal EMG alone achieved $43–46\%$ F1, demonstrating the discriminative strength of facial EMG as a complementary modality. Highly expressive emotions such as happiness and surprise were recognized most reliably (up to 0.84 and 0.75 true positive rates, respectively), whereas subtler expressions (anger, disgust) remained more challenging due to overlapping muscle activation patterns. 

\begin{figure*}[h!]
    \centering
    \includegraphics[width=0.9\textwidth]{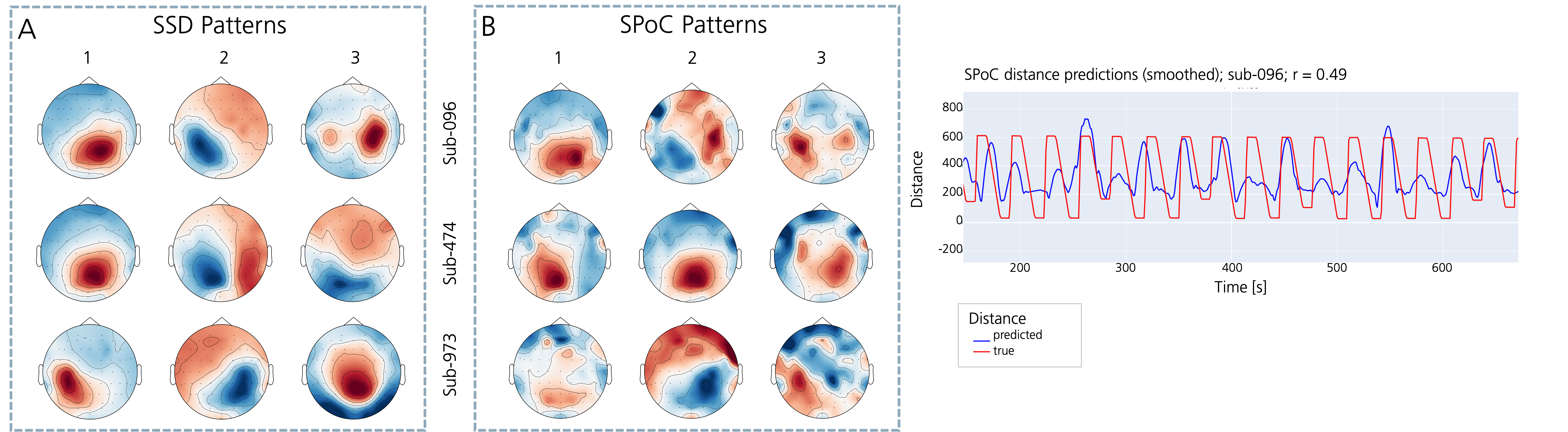}
	\caption{Activation patterns in 3 participants related to emotional arousal adapted from~\cite{nierula_2025_sfn}. }
    \label{fig:eegresults}
\end{figure*}

For \textbf{mental state decoding}, we present preliminary results from 3 participants. Figure~\ref{fig:eegresults}A shows the SSD-derived spatial filters, which successfully isolated alpha-band oscillations (8-12\,Hz). Subsequent SPoC analysis (Figure~\ref{fig:eegresults}B) identified neural sources whose alpha power fluctuations showed maximal covariation with changes in avatar distance. All participants exhibited SPoC patterns with prominent parieto-occipital activity, consistent with neural correlates of emotional arousal~\cite{hofmann2021_rollercoaster}. However, the ranking of SPoC components by correlation strength varied across participants. Understanding the optimal selection strategy of SPoC patterns is subject of ongoing analysis. The time course panel (Figure~\ref{fig:eegresults}C) illustrates decoding performance for one representative participant (sub-096): the predicted avatar distance reconstructed from SPoC-extracted alpha power (blue) tracks the actual avatar distance (red) with r=0.49.

\begin{figure*}[h!]
    \centering
    \includegraphics[width=\textwidth]{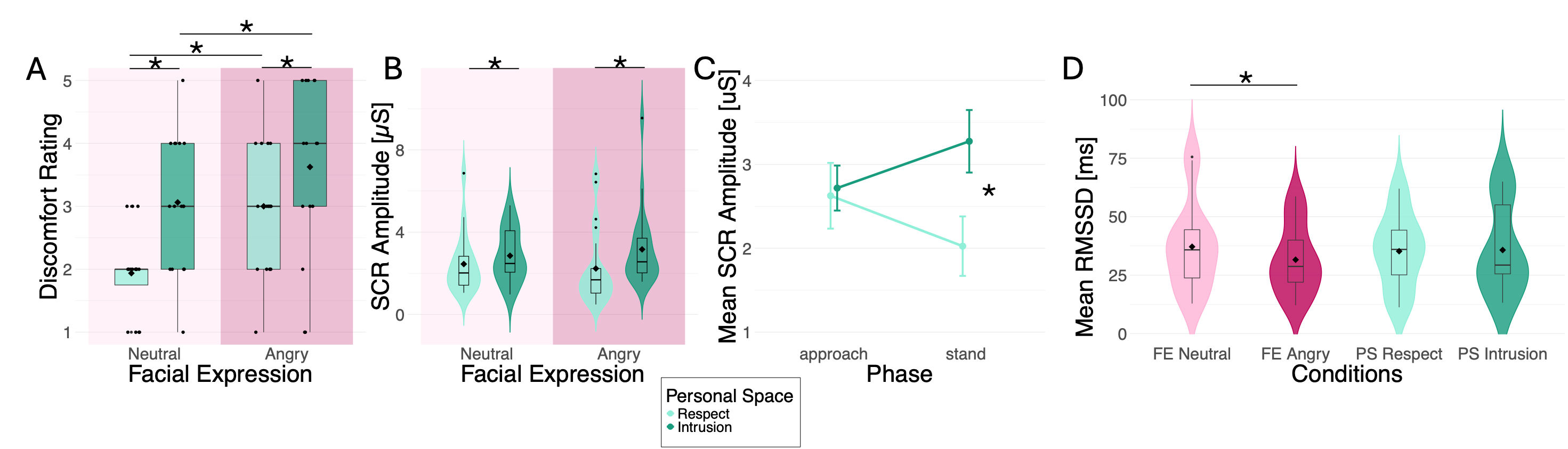}
	\caption{Physiological results adapted from~\cite{nierula_2025_physio}. (A) Discomfort ratings to an avatar that respected the user's personal space (light green) or violated it (dark green) with neutral or angry facial expression. (B) Effect on SCR amplitudes. There was only a main effect of Personal Space on SCR amplitudes. (C) Interaction effect of the avatar phase (whether approaching or standing in front of the user) on SCR amplitudes. (D) Effect on heart rate variability (HRV, measured via RMSSD). Asterisks in all panels indicate significant statistical effects (p$<$.05).}
    \label{fig:physioresults}
\end{figure*}

For \textbf{bodily arousal}, physiological markers of emotional arousal were tested in an experimental scenario where the avatar did or did not respect the user's personal space and showed a neutral or angry emotional facial expression~\cite{nierula_2025_physio}. Multimodal assessment revealed that our analysis methods were sufficiently sensitive to detect distinct autonomic nervous system responses across modalities. Skin conductance response (SCR) analysis proved sensitive to capture subtle changes in sympathetic arousal due to personal space violations (see Figure~\ref{fig:physioresults}B and C), while remaining unaffected by facial expressions. Conversely, heart rate variability (HRV) analysis demonstrated sufficient sensitivity to detect parasympathetic modulation in response to threatening facial cues, but showed no response to spatial violations (see Figure~\ref{fig:physioresults}D). 

Eye tracking evidence further demonstrates the experiential relevance of proxemic signals: interpersonal distance to a virtual instructor significantly affected visual attention patterns, with greater distance leading to attentional dispersion and reduced learning performance~\cite{lafci_2025}.

Collectively, these results establish the feasibility of automated non-verbal cue recognition for XR-based training applications and underscore the value of multi-view sensing and multimodal fusion strategies for overcoming occlusion and viewpoint challenges inherent to immersive environments.

\section{Discussion}
\label{sec:discussion}

This work presents an early-stage exploration of how multimodal, real-time communication analysis can be used to shape adaptive XR training experiences for de-escalation. Rather than framing multimodal sensing as an end in itself, we treat it as an interaction layer whose interpretation directly influences avatar behaviour and the trainee’s lived experience. Our discussion therefore focuses on design implications that arise when transforming heterogeneous communication signals into meaningful, experience-driven adaptations in XR.

A key insight is that multimodal signals only become meaningful through interpretation within a specific interactional context. Gestures, prosody, physiological arousal, and spatial behaviour do not map directly to communicative intent, but must be translated into experiential states such as perceived threat, calmness, or authority. In XR training, these mappings shape how virtual characters respond and whether interactions feel plausible and pedagogically useful. As a result, decisions about which cues matter and how they influence adaptation are fundamentally design choices, not purely technical ones.

Multimodal fusion emerged as a central and unresolved design challenge. The modalities considered in this work operate on different temporal scales, vary in reliability, and may produce conflicting signals. Static or globally optimized fusion strategies risk overreacting to transient signals or privileging modalities that are experientially misleading in a given context. As our fusion approach is still conceptual, we intentionally frame fusion as a design space rather than a finalized solution. Context-sensitive weighting, temporal smoothing, and conservative adaptation strategies appear particularly important for maintaining plausibility and user trust in immersive training scenarios. From an XR perspective, fusion should support experience shaping rather than maximize classification confidence.
Importantly, uncertainty in multimodal fusion is not only a technical limitation but also an ethical concern, as overconfident interpretations risk attributing intent, aggression, or communicative failure where none exists.

In the current implementation, each analysis stream operates independently, producing parallel feature outputs that are synchronized via LSL timestamps. While the architecture is designed to support multimodal fusion—where correlations, temporal dependencies, and potential conflicts between streams would inform unified interpretations—this integration remains conceptual at the present stage. Specifically, we have not yet implemented mechanisms for (1) weighting modalities based on context-dependent reliability, (2) resolving contradictory signals (e.g., relaxed posture but elevated physiological arousal), or (3) learning cross-modal temporal patterns that may be more predictive than any single stream.

We intentionally frame fusion as an open design space rather than a solved problem. Future work will explore both rule-based and data-driven fusion strategies, investigate how scenario context (encoded in the adaptive storybook) can dynamically adjust modality weights, and evaluate how different fusion approaches affect trainee experience and pedagogical outcomes. Until such strategies are validated, the system outputs per-stream features for human review rather than fully automated adaptation decisions.

Another important consideration concerns feedback and immersion. While real-time adaptation of avatar behaviour can provide powerful implicit feedback, overly frequent or explicit responses risk disrupting emotionally charged interactions. Our approach therefore favours embedding feedback primarily in avatar behaviour, complemented by post-scenario reflection and instructor-mediated interpretation. Deciding when not to adapt is as critical as deciding how to adapt, underscoring the importance of restraint and pacing in adaptive XR design.
Interpreting emotional, physiological, and behavioural signals in real time raises ethical challenges that directly shape system design. Physiological arousal or non-verbal behaviour cannot be unambiguously mapped to intent or escalation risk, and such signals are sensitive to individual, cultural, and situational factors. To mitigate the risk of over-interpretation, we intentionally favour conservative, context-sensitive adaptations and avoid treating system outputs as objective assessments. The system is designed to support reflection rather than evaluation, with trainers retaining responsibility for contextualizing feedback. This human-in-the-loop approach reflects both ethical considerations and the inherently ambiguous nature of interpersonal communication.

Given the sensitivity of de-escalation training, full automation is neither desirable nor appropriate. We position our system as decision-support rather than decision-making, with human oversight remaining central. Trainers contextualize feedback, and scenario authors define interpretation and fusion rules aligned with pedagogical goals. This highlights the need for XR authoring tools that make multimodal interpretation transparent and adjustable, enabling designers to shape system behaviour without relying on opaque black-box models. Ethical considerations such as privacy, consent, and the risk of misclassification further reinforce the importance of keeping trainees and instructors in control and framing feedback as formative rather than evaluative.

This work reflects an ongoing research effort with several limitations. The presented findings are preliminary and are primarily intended to inform design decisions rather than providing a comprehensive evaluation of training effectiveness or measuring training impact. Future work will focus on refining multimodal fusion strategies, exploring adaptive and personalized interpretations, and conducting larger-scale and longitudinal studies. While our case study centres on law enforcement, the design considerations discussed here are applicable to a wide range of XR training scenarios involving complex interpersonal interaction, including healthcare, education, and negotiation.


\begin{acknowledgments}

We would like to express our deepest gratitude to A. Bock, M. Heinze, D. Knuth, D. Martin, J. Schander, A. Orinsky, L. Braunisch, M. Ernst, D. Guthor, J. Gutman and our associated project partners Bayerische Polizei and Polizei Berlin for their valuable insight and guidance. This research was funded by the German Federal Ministry of Education and Research under the grants K3VR (13N16388).
\end{acknowledgments}

\section*{Declaration on Generative AI}
 During the preparation of this work, the authors used different commercial LLMs for grammar and spelling checking. After using these tools, the authors reviewed and edited the content as needed and take full responsibility for the publication’s content. 

\bibliography{citation}

\appendix

\end{document}